\documentclass[pra,twocolumn,aps,showpacs,10pt]{revtex4-1}
\usepackage[colorlinks,bookmarks=false,citecolor=blue,linkcolor=red,urlcolor=black]{hyperref}
\usepackage{graphicx}
\usepackage{amssymb,amsmath} 
\usepackage{amsfonts}
\usepackage{color}
\usepackage{braket}
\usepackage[utf8]{inputenc}
\usepackage[T1]{fontenc}
\usepackage{lmodern}
\usepackage{epsfig}
\usepackage{ulem}
 \newcommand{\da}{\downarrow}
\newcommand{\ua}{\uparrow}

 \newcommand{\be}{\begin{equation}}
 \newcommand{\ee}{\end{equation}}

\begin{document}

\title{Topological Rice-Mele model in an emergent lattice: Exact diagonalization approach}

\author{Krzysztof Biedro\'n$^1$, Omjyoti Dutta$^1$, and Jakub Zakrzewski$^{1,2}$}
\affiliation{\mbox{$^1$Instytut Fizyki imienia Mariana Smoluchowskiego, Uniwersytet Jagiello\'nski, \L{}ojasiewicza 11, 30-048 Krak\'ow, Poland}
\mbox{$^2$Mark Kac Complex Systems Research Center, Jagiellonian University, \L{}ojasiewicza 11, 30-348 Krak\'ow, Poland}
}

\begin{abstract}
  Using exact diagonalization methods we study possible phases in a one--dimensional model of two differently populated fermionic species in a 
  periodically driven optical lattice.  The shaking amplitude and frequency are chosen to resonantly drive $s-p$ transition
  while minimizing the standard intra-band tunnelings. We verify numerically the presence of an emergent  density wave configuration of composites for appropriate filling fraction and minimized intra-band tunnelings.
  The majority fermions moving in such a lattice mimic the celebrated Rice-Mele model.
  Far away from that region, the structure changes to a clustered phase, with the intermediate phase abundantly populated by defects of the density wave. These defects lead to localized modes carrying fractional particle charge. 
The results obtained are compared with earlier approximate predictions.
\end{abstract}

\maketitle

\section{Introduction}
Ultracold atoms trapped in optical lattices provide systems characterized by an unprecedented control over various parameters, enabling a simulation of a wide array of exotic solid state models. One example of such phenomena are topological insulators\cite{Hasan10,Liang11} that are of particular interest in the field of quantum information and spintronics due to their inherent stability and transport properties \cite{Nayak08,Moore2010,Pesin2012}.
 Lattices hosting systems showing topological properties have been realized experimentally, both for two--dimensional (2D)\cite{Tarruell2012,Hauke12,Chang12042013,Aidelsburger13,Jotzu2014,Aidelsburger2015} and one--dimensional (1D) models (e.g. SSH/Rice-Mele dimer \cite{SSH,RM} in \cite{Atala2013} or Thouless pump in \cite{lohse2015thouless,nakajima2015topological}). Optical lattices do not allow by themselves for
generation of impurities on which boundary localized modes may appear -- the lattices are necessarily perfect. 
In 2D, the localized defect -- a vortex -- may be created by a vortex wave \cite{Schonbrun06,Kartashov06} leading to a well placed dislocation. In 1D, the situation is not so simple, but a recent proposition \cite{Dutta14}
suggests that topologically nontrivial states may emerge in systems consisting of two subspecies of strongly attracting fermions. There, the topological structure is not encoded in the underlying lattice geometry, but rather is an emergent feature arising from atomic interactions, enabling creation of  defects with less constraints. For high enough values of the  interaction strength, fermions of different species tend to bind together forming composites, and if there is some imbalance in a number of atoms of both species, excess fermions stay unbound. To extract essential properties of the system, one has to take into account higher bands ($p$ band at least, as in the model studied below) and the effects of strong interactions, such as the density induced tunnelings\cite{pdz15,Mering11,Luhmann12,Duttarev}. The lattice shaking  is employed with the shaking frequency such that the inter-band density-induced $s$ to $p$ tunneling is resonantly enhanced.  As a result, in a 1D chain, the emergent system is 
proposed to 
be described by the
Rice-Mele model \cite{pdz15}. For a triangular lattice geometry, similar processes lead to the creation of  synthetic non-Abelian fields in an emergent dice lattice \cite{Dutta2015}. 

Let us note, parenthetically, that physics of $p$-orbital fermions is very rich, leading to a possible creation of FFLO states
\cite{Cai11} as well as density stripes at appropriate fillings due to nested Fermi surfaces \cite{Zhao08,Zhang12} even in the absence of any periodic driving (for a review of these effects see \cite{Liliu15}).
Those systems were studied using both two- and three-dimensional models. Here, we shall restrict ourselves to small 1D systems amenable to exact diagonalization.
\begin{figure}
\includegraphics[width=60mm]{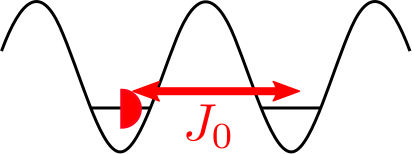}
\includegraphics[width=60mm]{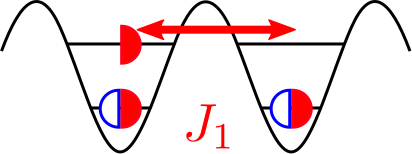}
\caption{ \label{tun}{Visualization of the kinetic tunneling processes present in the system. Blue (open) and red (filled) half-circles denote the minority, $\ua$-fermions and the majority, $\da$-fermions, respectively. 
Observe that the minority fermions appear paired in composites only.}}
\end{figure}

Let us stress that  the main approximation used in periodically driven models discussed in \cite{pdz15, Dutta2015} is to neglect the tunneling of the minority components. 
As a result, one generates a modified Falicov-Kimball like model with immobile composites (made out of strongly coupled pair of fermions) and mobile excess fermions.
We test this assumption in the present paper. Namely, we are  employing an exact diagonalization method to
the system described in \cite{pdz15} in order to assess the validity of the results presented there. The fidelity and structure factor analysis allow us to classify the ground states for different values of parameters.  We consider also explicitly possible configurations with a given number of defects.

\section{The system}

The system considered is a mixture of two species of unequally populated, strongly attractively interacting fermions in a 1D periodically shaken optical lattice. The Hamiltonian of the system is \cite{pdz15}
 $\hat{H}=\hat{H}_{\rm tun}+\hat{H}_{\rm dit}+\hat{H}_{\rm ons}+\hat{H}_{\rm sh}(t),$ where: 
\begin{eqnarray}
\label{Ht}
\hat{H}_{\rm tun}&=&- J_0\sum_{ \langle {ij}\rangle}  \left [\hat{s}^\dagger_{i}\hat{s}_{j}+ \hat{s_\ua}^\dagger_{i}\hat{s_\ua}_{j}\right]+ J_1 \sum_{ \langle {ij}\rangle} \hat{p}^\dagger_{i}\hat{p}_{j},\nonumber\\
\hat{H}_{\rm dit}&=&
\sum_{\langle {ij}\rangle}\left [ T_0 \hat{s_\ua}^\dagger_i(\hat{n}^{}_ {i}
+\hat{n}^{}_ {j})\hat{s_\ua}_{j}+T_1{\hat p_{i}}^{\dagger}(\hat{n}^\ua_ {i}
+\hat{n}^\ua_ {j})     \hat{p}_{j}
 \right. \nonumber\\
&\quad&+\left. 
 T_{01}((j-i){\hat p_{{i}}}^{\dagger}\hat{n}^\ua_ {i} \hat s_{{j}}^{} + h.c )  \right ],\\
\hat{H}_{\rm ons}&=& U_{0}^{}\sum_{i}\hat{n}^\ua_i \hat{n}_i+ U_{1}^{}\sum_{i}\hat{p}^\dagger_i\hat{p}_i\hat{n}^\ua_i+E_1\sum_{i}\hat{p}^\dagger_i\hat{p}_i, \nonumber \\
\hat{H}_{\rm sh}(t)&=& K \cos \omega t \sum_{j}  j  (  
\hat{n}^\ua_j + \hat{s}^\dagger_j\hat{s}_j+
\hat{p}^{\dagger}_j\hat{p}_j ) \nonumber \\
&&+  \delta E_1\cos (\omega t +\varphi) \sum_{i}{\hat p_{ {i}}}^{\dagger}\hat{p}_{{ {i}}}^{}.\nonumber
\end{eqnarray}
Above and in the following, 
$\hat{s}^{\dagger}_{{i}}, \hat{s}_{{i}}^{}$, $\hat{p}^{\dagger}_{ {i}}, \hat{p}^{}_{ {i}}$ are creation and annihilation operators of $\downarrow$-fermions in the $s$- and $p$-bands respectively, while $\hat{s}^{\dagger}_{\ua{i}}, \hat{s}_{\ua{i}}^{}$ are $s$-band creation and annihilation operators for $\ua$-fermions.  Accordingly, $\hat n_i, \hat n^p_i,$ and $\hat n^\ua_i$ are the corresponding
number operators. Note that while we take into account $s$ and $p$ bands for $\da$-fermions we consider only $s$ band  for
$\ua$-component. That is so because we assume that $\ua$-fermions form a minority component with filling close to 1/2. On the other hand, we assume a bigger density for $\da$-fermions.
\begin{figure}
\includegraphics[width=60mm]{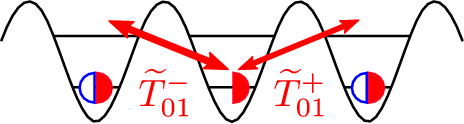}
\caption{ \label{tun2}{Visualization of the direction dependent inter-band density dependent  tunnelings in the dressed picture.
The dressed tunneling amplitudes are direction dependent.}}
\end{figure}

The single-particle tunneling part of the Hamiltonian is given by $\hat{H}_{\rm tun}$. We assume both species to have the same mass and feel the same optical lattice for simplicity. With the adopted sign convention $J_0,J_1>0$. The density dependent tunneling part is denoted as $\hat{H}_{\rm dit}$. The tunneling coefficients $T_0,T_1,T_{01}$ are given by appropriate integrals of Wannier functions \cite{pdz15,Duttarev}. Since the $p$-Wannier orbital is  antisymmetric, the inter-orbital tunneling amplitudes have opposite signs in opposite directions as reflected by $(j-i)$ factor.

The basic assumption of the model is that attraction between different species dominates  the problem energetically. Consider the onsite energy term $\hat{H}_{\rm ons}$. Under our assumption, $U_0$ is negative with $|U_0|$ giving the large energy scale.  $E_1$ -- the energy of the $p$ band -- is another large energy.
As tested by us with Wannier functions for different lattice depths,  $|U_1|$, the energy of the interaction between 
a fermion in the $p$ and in the $s$ band, is smaller than $|U_0|$. 

With that assumption, the lowest energy manifold is filled with composites -- pairs of $\ua$ and $\da$ fermions -- and the
remaining $\da$ fermions, leading to nontrivial dynamics. Note that, for example,  if  a minority $\ua$-fermion tunnels from a
given site, it leads to breaking of the composite. It costs a huge amount of energy ($|U_0|$) unless the tunneling occurs to a
site in which a majority $\da$-fermion waits to form a composite with the $\ua$-particle. Only the latter process remains in the low energy manifold. In effect, the simple tunneling of the minority fermion may be viewed as a tunneling of the
composite, accompanied by a reverse direction tunneling of the majority fermion (compare Fig.~\ref{tun2}) within this manifold.  The system may be described by operators describing excess majority fermions (residing either in $s$ or in $p$ band) and the
composites described by annihilation (creation) operators $\hat{c}_i $($\hat{c}_i^\dagger$) obeying hard-core boson commutation relations. The corresponding 
composite number operator is  $\hat{n}^c_i=\hat{c}_i^\dagger\hat{c}_i$. The presented intuitive picture is fully recovered on a more formal level by an appropriate construction of the effective Hamiltonian \cite{Dutta14}. In effect, the excess majority fermions move in the emergent lattice created by the composites. To avoid excessive repetitions we refer the reader to \cite{Dutta14} for details
while \cite{Dutta2015} provides yet another example of a two-dimensional construction based on the idea described above.

The second important step is to derive the effective Hamiltonian valid for the high frequency driving obeying
the (almost)  resonant condition 
\be
U_1+E_1=N\omega+\Delta
\label{rez}
\ee
with $N$ being an integer and a small detuning $|\Delta|\ll \omega$. Observe that the time--dependent part of the Hamiltonian, $\hat{H}_{\rm sh}(t)$, contains two time--periodic terms. The first one describes a standard horizontal lattice shaking (after an appropriate gauge transformation) as originally proposed in \cite{Eckardt05} and reviewed e.g. in \cite{Arimondo12}. Such a horizontal shaking has been realized experimentally by several groups \cite{Lignier07,Zenesini09,Struck11} and serves as a convenient knob on lattice system properties.
The second term is due to the harmonic variation of the lattice depth. This translates into a periodic modulation of the $p$-band energy offset $\delta E_1$ \cite{pdz15}. The phase $\varphi$ between the two harmonic modulations can be easily controlled in experiments. The procedure of averaging is fairly standard and is described in detail in \cite{pdz15}. We quote here the final effective Hamiltonian expressed in terms of composite and excess fermion operators:

\begin{align} \label{hammin}
  \begin{split}
    \hat{H} = & \widetilde{T}_{01}^- \sum_{i=1}^N (\hat{p}_i^{\dagger} \hat{n}^c_{i} \hat{s}_{i+1} + \textrm{h.c.}) \\
              & - \widetilde{T}_{01}^+ \sum_{i=1}^N (\hat{p}_i^{\dagger} \hat{n}^c_{i} \hat{s}_{i-1} + \textrm{h.c.}) \\
              & + (2\widetilde{T}_0 - \widetilde{J}_0) \sum_{i=1}^N (\hat{c}_i^{\dagger}\hat{s}_{i+1}^{\dagger}\hat{c}_{i+1} \hat{s}_i +\textrm{h.c.}) \\
              & + (2\widetilde{T}_1 + \widetilde{J}_1) \sum_{i=1}^N (\hat{p}_i^{\dagger} \hat{n}^c_{i} \hat{n}^c_{i+1} \hat{p}_{i+1} + \textrm{h.c}) \\
              & - \widetilde{J}_0 \sum_{i=1}^N \hat{s}_i^{\dagger} \hat{s}_{i+1} + \Delta \sum_{i=1}^N \hat{n}^p_{i},
  \end{split}
\end{align}

where the tilde sign over tunnelings and density dependent tunnelings indicates their dressed character (after time averaging). Explicitly, $\widetilde{J}_l =\mathcal{J}_0\left(\frac{K}{\omega}\right)J_{l}$ with $l=0,1$ corresponding to $s$ and $p$ band respectively and $\mathcal{J}_0$ being the Bessel function \cite{Eckardt05,pdz15}.
 A  similar dressing  
  takes place for intra-band density dependent tunnelings $T_l$.  On the other hand, 
  the inter-band $s-p$ tunneling amplitude value becomes  direction dependent  due to the phase difference between shaking amplitudes. We express that asymmetry by denoting the tunnelings between  $p_i \leftrightarrow s_{i+1} $ as $  \widetilde{T}^{+}_{01}$ and $p_i \leftrightarrow s_{i-1}$ as $ \widetilde{T}^{-}_{01}$. 
  These tunneling processes are visualized in Fig.~\ref{tun2} and read explicitly \cite{pdz15} $\tilde T_{01}^\pm =\mathcal{J}_N\left( A^\pm/\omega\right)T_{01},$ where $A^\pm =\sqrt{(K\pm\delta E_1 \cos\varphi)^2+K^2 \sin^2\varphi}$.

While the frequency of the periodic drive is fixed by the resonance condition \eqref{rez}, the shaking amplitude $K$ provides a convenient parameter to tune the properties of the system. In particular,  $K_c$
such that  $K_c/\omega\approx 2.405$ corresponds to the zero of $\mathcal{J}_0$ Bessel function. For such a choice of $K,$
the intra--band tunnelings almost vanish and the inter--band density dependent tunneling becomes the only mechanism of transferring the majority fermions (the composites becoming immobile in this limit).   Then, as suggested in \cite{pdz15} for $n^\ua=1/2$ (the filling for minority fermions), the composites form a density wave (DW) in the ground state while the excess majority fermions are described by a Rice-Mele topological dimer model. On the other hand, sufficiently far from $K_c$ the standard tunneling mechanisms dominate -- the system then organizes into a clustered phase (CL) with composites and empty sites separated in space \cite{pdz15}. 

To test this prediction, one has to carefully estimate various parameters appearing in the minimal Hamiltonian, \eqref{hammin}. They depend on the details of the
lattice potential and interactions between two species. We follow the assumptions of \cite{pdz15}
and assume the optical lattice potential to take the form $V_{\mathrm{latt}}= V_{\parallel}\sin^2(\pi x/a) + V_{\perp}(\sin^2(\pi y/a) + \sin^2(\pi z/a)) $, with $a$ being the lattice constant. For $V_{\perp}\gg V_{\parallel}$  the system is effectively one--dimensional. We take $V_{\perp}=25$ while $V_{\parallel}=8$ in the units of the recoil energy  $E_R=h^2/(8Ma^2)$ (note that $a=\lambda/2$ with $\lambda$ being the wavelength of the laser beams forming a standing wave pattern). As a dimensionless interaction strength we take a plausible value $\alpha=a_s/a=-0.1$ (with $a_s$ being the (negative) scattering length). That, together with lattice parameters, allows us to estimate all the tunneling and interaction parameters of the model using the Wannier functions appropriate for the lattice potential \cite{pdz15}. 

\begin{figure}
  \centering
  \includegraphics[width=0.9\columnwidth]{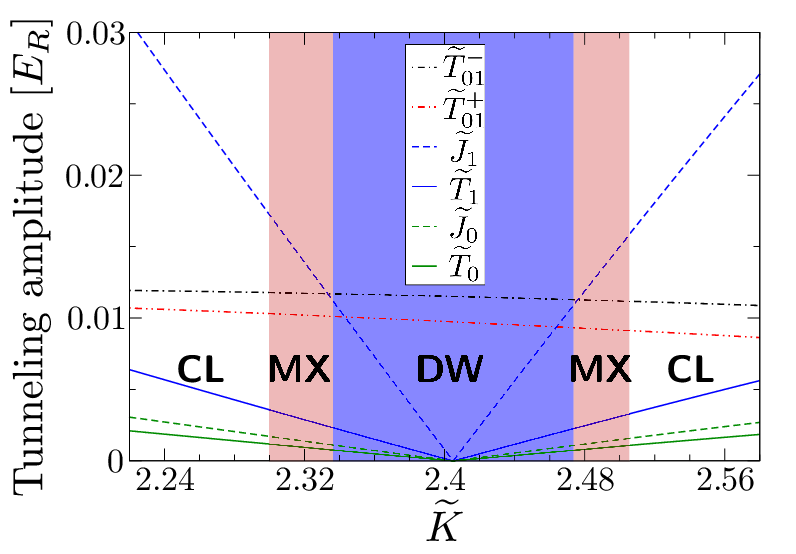}
  \caption{$\widetilde{J}_i$, $\widetilde{T}_i$ and $\widetilde{T}_{01}^{\pm}$ tunnelings for $V_0 = 8$, $V_{\perp} = 25$, $\alpha = -0.1$, $N=1$ and $\phi = 0$. For $\tilde K=K/\omega$ range shown in the figure there exist three phases: the density wave (DW) (blue), the clustered phase, CL (white) and the mixed phase, MX (pink). Boundaries of these phases were calculated using the fidelity susceptibility (see fig \ref{fig:fidelity}) and the structure factor (fig.~\ref{fig:sfactor}).}
  \label{fig:tunph}
\end{figure}

As far as the shaking is concerned, we obviously concentrate on the vicinity of $K/\omega=2.4$ region, taking the vertical shaking  to be in phase with the lateral one  ($\varphi=0$), which  gives ${\tilde T^-_{01}> T^+_{01}}$. For simplicity, we assume first the exact driving resonance $\Delta=0$. 
 In Fig. \ref{fig:tunph} we show the dependencies of the different dressed tunnelings as a function of $K/\omega$
 (we shall later assume a notation $\tilde K=K/\omega$) coming from Wannier function calculations.

To find the ground state of \eqref{hammin} we have yet to define the density of majority component which is taken to be unity (thus we have a 1/2-filling of composites and 1/2 filling of excess fermions). Then, we
use the  exact diagonalization method based on \cite{sandvik2011computational,0143-0807-31-3-016}. Diagonalizations take place in Fock space of all possible configurations of the system, assuming that each site $i$ may be empty or occupied by a single composite or a $\downarrow$ fermion in $s$ state, or both the composite and $\downarrow$ fermion, although the second one in $p$ state (because there is already $s$-state $\downarrow$ fermion in a composite). Therefore, the local Hilbert space consists of 4 states per site with no truncation.  
For an even number of fermions, a fermion tunneling between arbitrary ``edges'' (that is, between the first and the last site) leads to an additional phase (sign) change arising from the anti-commutation relations. 
Because the number of fermions is half the number of sites, available numbers of sites are of the form of $L=4l+2$, $l\in\mathbb{Z}$.

With periodic boundary conditions, Hamiltonian in Eq.\eqref{hammin} commutes with the translation operator ($\hat{T}$), which allows us to use states with the conserved total momentum ($k$) as our basis: $T\ket{a(k)} = e^{i k}\ket{a(k)}$. States with different $k$s are orthogonal to each other, and $k\in\{(-L/2+1)\frac{2\pi}{L}, \ldots, (L/2)\frac{2\pi}{L}\}$ (because $T^L\ket{a(k)} = e^{i k L}\ket{a(k)} =\ket{a(k)}$) with $L$ being the length of the chain. Diagonalization consists of creating states in the basis (for some/all values of $k$), calculating matrix elements of $\hat{H}$ in that basis and using numerical algorithm to get eigenvalues and eigenvectors for the lowest energy states. We would like to point out that the total momentum $k$ serves only  to split the large Hamiltonian matrix into smaller blocks.  

\section{Results}

We carry out exact diagonalizations typically on a chain of length $L=14$ (leading to matrices of the rank $\sim 840 000$). For selected data we show the results for $L=18$ (matrices of rank around 131 million).  As tunnelings are nearly symmetric with respect to $\tilde K_c \approx2.405$ (only $T_{01}^{\pm}$ are noticeably different, which leads to small, quantitative -- but no qualitative -- changes), we will only consider $\tilde K  < 2.405$.
In the interval of interest, the ground state corresponds to $k = \pi$. To characterize its properties and locate possible phase transitions we use the fidelity approach \cite{Zanardi06}. 
We calculate the fidelity, $F$, associated with a small parameter change $\delta$ (here $F(\tilde K, \delta) = \braket{\psi_0(\tilde K-\delta/2) | \psi_0(\tilde K+\delta/2)}$) using the eigenvectors coming from the diagonalization. For $\delta\approx 0$, we get $F\approx 1 - \chi \frac{\delta^2}{2}$ which defines the fidelity susceptibility, $\chi$  \cite{PhysRevE.76.022101,Damski13}. It is commonly understood that the fidelity susceptibility diverges at phase transitions. For our finite system, the possible crossovers will be identified by the maxima of $\chi$.

\begin{figure}
  \centering
  \includegraphics[width=0.9\columnwidth]{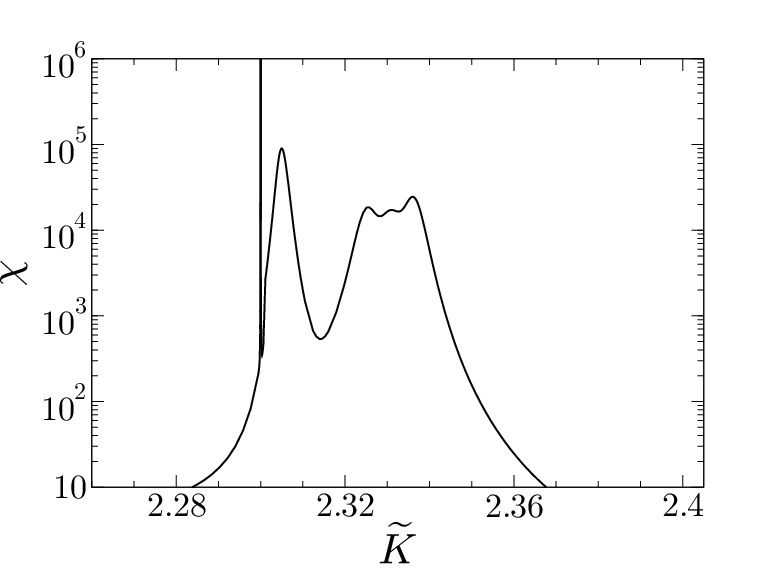}
  \caption{Fidelity susceptibility, $\chi$, for the ground state for $L=14$. Observe a sharp peak at $\tilde K=2.3$ and a rich structure of peaks up to about $\tilde K=2.35$. The different phases are
  analyzed using the structure factor -- see the text and fig.~\ref{fig:sfactor}}
  \label{fig:fidelity}
\end{figure}

\subsection{Resonant case}

As mentioned above, the simplified analysis \cite{pdz15} predicts the existence of two composite arrangements: the density wave (DW) close to $\tilde K_c=2.405$, where intra--band tunnelings are effectively switched off, and the clustered phase (CL), where the composites group together. Thus, we should expect a single $\chi$ maximum for $\tilde K<\tilde K_c$ corresponding to the border between these two phases. The numerical results are,
however, quite different (compare Fig.~\ref{fig:fidelity}). There are indeed two regions of low fidelity susceptibility for $\tilde K<2.29$ (with a sharp peak of $\chi$ around $\tilde K=2.3$) as well as for $\tilde K$ values close to $\tilde K_c$ (for $\tilde K \gtrsim 2.35$) indicating stable phases. On the other hand, the interval $\tilde K \in (2.29,2.35)$ shows a structure of peaks with $\chi$ having significant values almost everywhere.

\begin{figure}
  \centering
  \includegraphics[width=0.9\columnwidth]{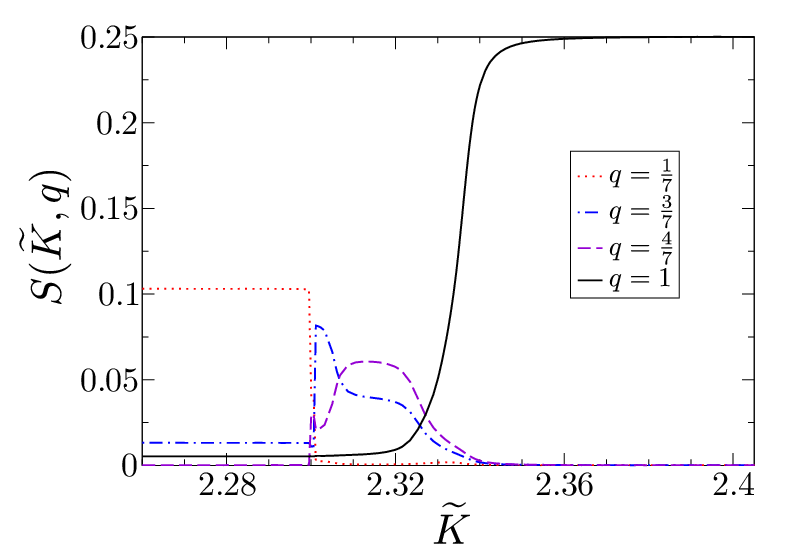}
  \caption{Structure factor $S(\tilde K, q)$  calculated for $L=14$ sites and different $q$ values as indicated in the figure.  For $\tilde K>2.35$ a single $q=1$ value dominates indicating DW phase. Similarly For $\tilde K<2.29$ $q=1/7$ dominates pointing out to the clustered phase (CL). Changes in the structure factor behavior nicely correlate with peaks in the fidelity susceptibility -- compare fig.~\ref{fig:fidelity}).}
  \label{fig:sfactor}
\end{figure}

To understand that somewhat complicated behavior of $\chi$ we consider   the structure factor here defined as
\begin{equation}
  S(\tilde K, q) = \frac{1}{L}\left\langle\sum_{|i-j|}(\hat{n}_i^c - \frac{1}{2})(\hat{n}_j^c - \frac{1}{2}) e^{-iq\pi|i-j|}\right\rangle,
  \label{sfactor}
\end{equation}
where $\hat{n}_i^c$ is the number of composite bosons (0 or 1 in our case) occupying $i$-th site. 
Fig. \ref{fig:sfactor} shows three areas with different behavior of structure factor, each corresponding to different phase structure. For the CL phase, the structure factor $S(\tilde K, q=2/L)\approx 0.1$, while values for different $q$ are close to 0, which happens to be the case for $\tilde K$ sufficiently far from $\tilde K_c$. On the other hand, for DW, $S(\tilde K, q=1) = 0.25$ and vanishes for other $q$ values. Such a behavior is seen close to the resonance, $\tilde K>2.35$. Thus indeed the two phases obtained close to the resonance and far from it show the properties predicted in \cite{pdz15}. Note that since the number of particles is strictly conserved in exact diagonalization, we cannot use some mean field order parameter to classify the phases observed. Still the identification based on the structure factor is unambiguous.

The behavior is more complicated in the intermediate interval of $\tilde K$ values. The structure factor for both $q=2/L$ and $q=1$ becomes small while intermediate $q$ values ($4/L, 6/L,..$) become important.
The situation seems somehow clearer close to the border of phase transitions.
  Around $\tilde K=2.3$ the peak in fidelity susceptibility coincides with the change in ground state structure (as seen in $S(\tilde K, q)$ plot) -- instead of the fully separated phases of composites and empty sites we observe splitting of the composites cluster into two (in small ($\Delta \tilde K \approx 0.001$) interval directly above $\tilde K = 2.3$) and three clusters (which corresponds to the dominant $S(\tilde K, 3/7)$ value). 
  Let us denote the pure CL phase as a string $\mathrm{0000000CCCCCCC}$ with C sites being filled by composites.
  Respective many-cluster phases can be traced back to $\mathrm{00000CC00CCCCC}$ and $\mathrm{000CC00CC00CCC}$ configurations as verified by a careful examination of the ground state wave-function expansion in Fock space (possible due to the small size of our system).
  On the other hand while in the vicinity of $\tilde K_c$ we observe a pure DW phase, close to $\tilde K=2.35$ the inspection of the wave-function reveals an addition of defected component, with 2 sites breaking the DW symmetry. The relative importance of a single defect component changes smoothly from practically zero close to $\tilde K_c$ (observe that above $K=2.35$ all $q$-components of $S(\tilde K, q)$ vanish except $q=1$) to become significant below $\tilde K=2.35$. The subsequent peaks of the fidelity susceptibility, $\chi$ in Fig.~\ref{fig:fidelity} nicely coincide with different components of  $S(\tilde K, q)$ dominating the structure factor. That corresponds, as again confirmed by the inspection of the wave-function components,
to successive defects of the partial DW leading to small clusters eventually merging as $\tilde K$ moves further away from $\tilde K_c$.
\begin{figure}
  \centering
  \includegraphics[width=0.9\columnwidth]{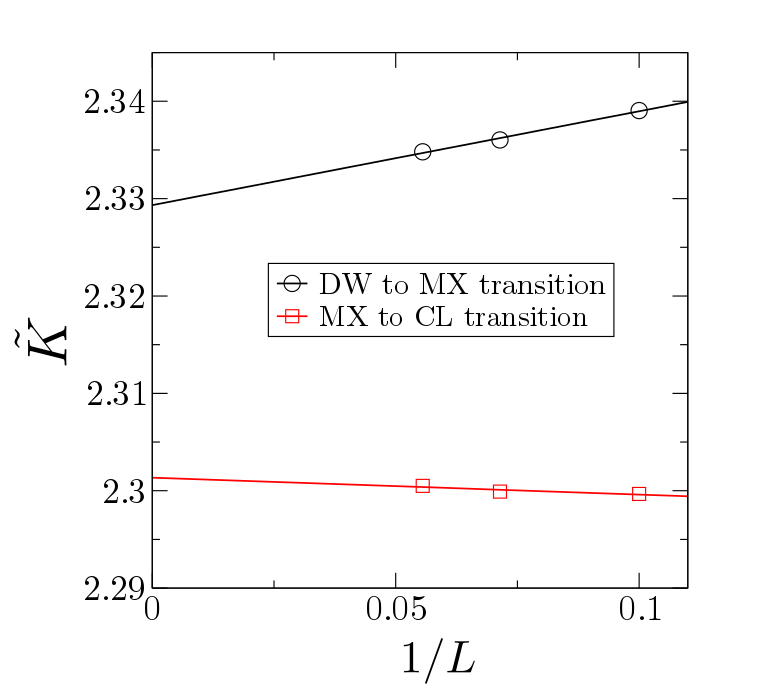}
  \caption{Borders between DW and MX phase as well as MX and CL phase as a function of the inverse system size. Observe that 
  the mixed phase persists in the extrapolated $L\rightarrow\infty$ limit. }
  \label{fig:trans}
\end{figure}

One may pose an important question whether the mixed phase observed is not really a finite size effect which will disappear
in the thermodynamic limit and the mean field analysis \cite{pdz15} will be recovered in that limit. To provide an answer,
we have evaluated the borders between different phases for the longer chain with $L=18$. Plotting the borders as a function of $1/L$ and extrapolating to the infinite chain one can see a clear indication that the mixed phase should not be purely a finite size effect. Let us note that this behavior is reminiscent of striped phases observed in two-dimensional Falicov-Kimball model \cite{Lemanski02}.  Importantly,
considering the standard optical lattices systems, the typical lattice size is about 50 sites thus the results obtained here are of a direct experimental relevance.

\begin{figure}
  \centering
  \includegraphics[width=0.9\columnwidth]{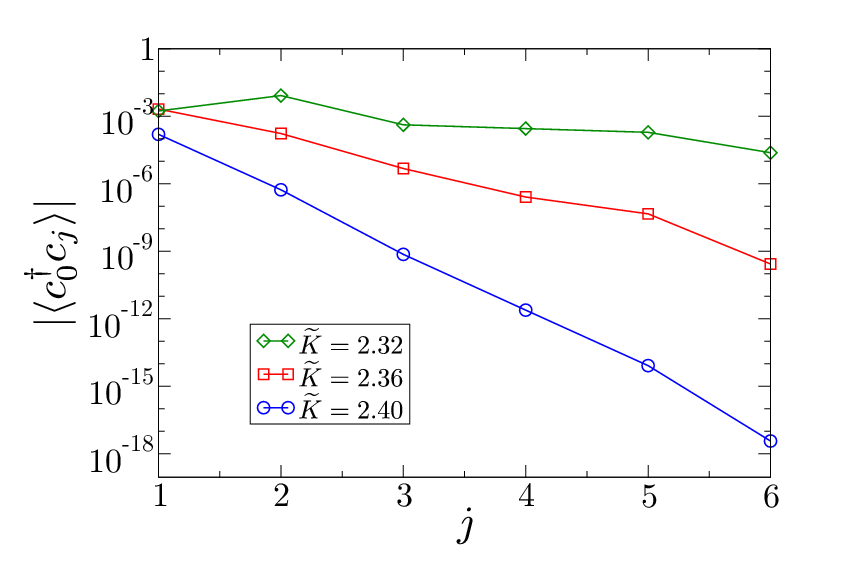}
  \caption{Correlation function $\langle c_0^{\dagger} c_j\rangle$ of creation operators for boson composites in ground state calculated for $L=14$ sites system for three values of parameter $\tilde K= K/\omega$, which are in the mixed 
  ($\tilde K = 2.32$) and density wave ($\tilde K= 2.36, 2.40$) phases.
  The correlation function decays exponentially in the DW phase, for the MX and CL (not shown) phase the decay is much slower, presumably power-like.}
  \label{fig:cij}
\end{figure}

Using results from diagonalizations, one can easily calculate the correlation function of composite bosons, $\langle c_0^{\dagger} c_j \rangle$.
When a system is in DW phase,  the correlation function decays exponentially with increasing $j$ 
$|\langle c_0^{\dagger} c_j \rangle|\propto \exp(-j/l_c)$ -- compare Fig.~\ref{fig:cij}. The correlation length $l_c$ depends strongly on $\tilde K$ -- compare the correlation functions  for $\tilde K=2.36$ and $\tilde K=2.40$. For other phases much slower decay, presumably power-like, is observed but no definite conclusions may be drawn due to small sizes considered. To that end,  one should perform a numerical study of a much larger chain, e.g. using density matrix renormalization group (DMRG)
which is beyond the scope of the present work.

\begin{figure}
  \centering
  \includegraphics[width=0.9\columnwidth]{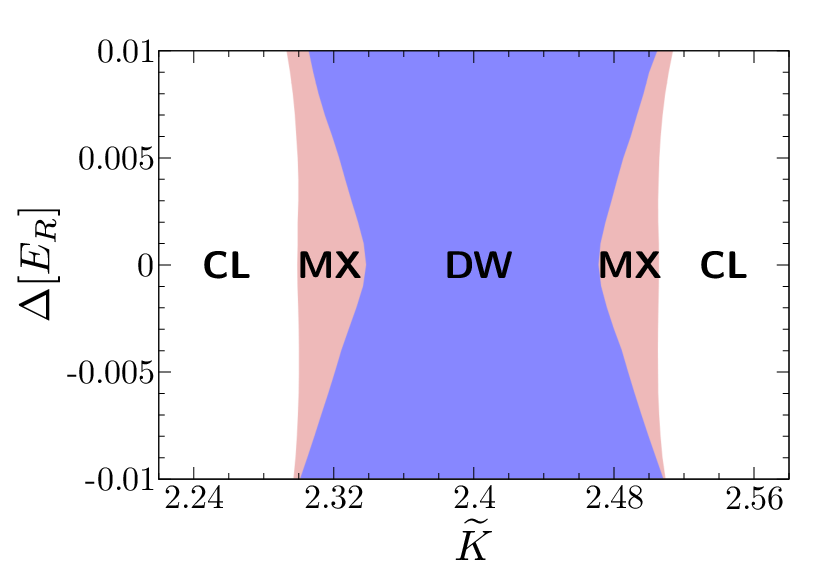}
  \caption{Different phases in the $\tilde K$ -- $\Delta$ plane. Observe the shrinking of the intermediate mixed phases region while the interesting density wave arrangement of composites region gets larger. The borders obtained for this plot were collected from diagonalizations of $L=10$ system with 5 composites, results for $L=14$ at selected points show that the picture is not affected significantly by the system size.}
  \label{fig:delta}
\end{figure}

\subsection{Detuned case}

While the studies of \cite{pdz15} and the results presented above concentrated on the $\Delta=0$ case corresponding  in DW phase to SSH Hamiltonian \cite{SSH}, it is interesting to see whether the full Rice-Mele model \cite{RM} for $\Delta\ne0$ leads to similar conclusions. To that end,
we have studied the phase diagram in the $\tilde K - \Delta$ plane as shown in Fig.~\ref{fig:delta}. Observe that while the border of the CL phase is not sensitive to $\Delta$, the region of DW actually increases 
eating up the MX phase. Therefore, the Rice-Mele model seems to be realized quite easily with the present system.

\begin{figure}
  \centering
  \includegraphics[width=0.9\columnwidth]{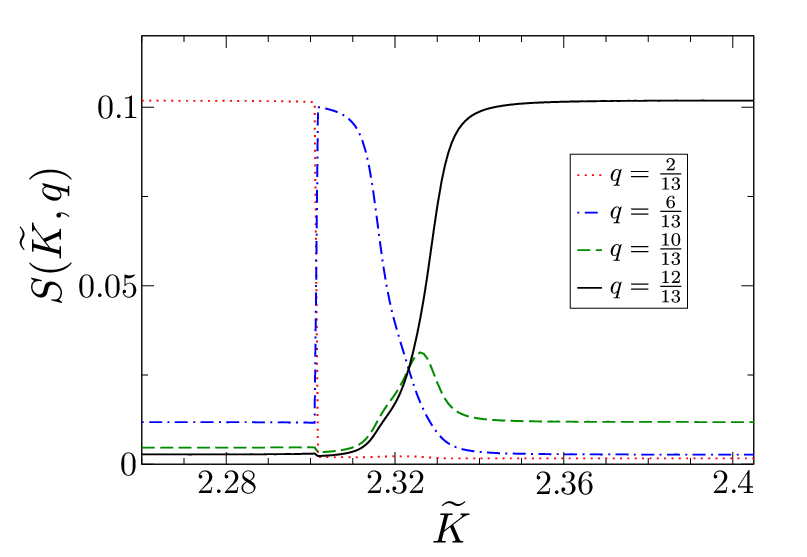}
  \caption{Structure factor $S(\tilde K,q)$ for $L=13$ and $N=6$ composites. Such a situation results in a defect (hole) present in the system. A single hole leads to a structure factor being dominant for $q=12/13$, the presence of a hole results in non vanishing values of $S$ for other $q$'s. As before, the CL phase is characterized by the structure factor being most prominent at $q=2/L$.}
  \label{fig:defect}
\end{figure}

The most interesting physics of the Rice-Mele model comes from localized modes on defects on the borders between topological and trivial phases \cite{RM, HKSS}. As discussed in \cite{pdz15}, the present model allows for control of the number of defects by changing slightly the filling of minority fermions, i.e. of composites. fFr the filling $n_c<1/2$
one creates holes in the DW, for $n_c>1/2$ we should have extra particles.   Indeed, as visualized in Fig.~\ref{fig:defect} when we consider 6 composites in $L=13$ sites, the DW phase (occurring for 7 composites and $L=14$ sites) is replaced by a ``single hole phase'' (SHP). Due to the periodic boundary conditions and the translational invariance of the system, the ground state is a combination of states with a hole at different positions along the lattice as revealed by the eigenstate inspection in the Fock representation. The border between a SHP and mixed configurations
is placed close to the value for the border of the DW phase in an ideal half filling of composites (taking into account finite size effects). For $\tilde K$ far from $\tilde K_c$, we observe a sharp phase transition to clustered phase with holes and defects separated.

\section{Conclusions}

Using exact diagonalization on small systems, we have addressed the problem of resonantly shaken optical lattices in which an unevenly populated mixture of two species of fermions is held. We have verified the basic model studied in \cite{pdz15} where, neglecting minority fermion tunnelings,  a density wave arrangements of composites were found in the situation when the shaking amplitude was tuned in a way enabling switching off all of the intra-band tunnelings. Then, the excess majority fermions move in an emergent lattice (formed by composites) with direction dependent tunnelings realizing the topological Rice-Mele model. In the  simplified approach \cite{pdz15}, it was found that apart from the density wave (for switched off intra-band tunnelings) the composites and empty sites may separate forming two clusters -- when the intra-band tunnelings are important. 
That has also been confirmed  by the present calculation. In addition to these two phases, the middle region separating these ideal 
case is revealed by an 
exact diagonalization.  In this mixed-phase region, the ground state contains superposition of many different composite arrangements. This phase may show quasi-long range order which is absent in the density wave phase.

We have also shown that the density wave phase in the vicinity of shaking parameters combination switching off intra-band tunnelings ($K/\omega\approx 2.405$ -- the zero of 0-order Bessel function) persists even when the shaking frequency is not adapted precisely to the $s$-$p$ orbital resonance condition -- thus it is quite robust. We have explicitly shown that the deviations from the ideal half filling of the minority fermions (and thus the composites) leads directly to defects (holes or extra particle) that, if occurring on the edge of the topologically nontrivial phase lead to localized modes.

\acknowledgments

This work was realized under National Science Center (Poland) project No. DEC-2012/04/A/ST2/00088. A support of EU Horizon-2020 QUIC 641122 FET program is also acknowledged.

\bibliographystyle{apsrev4-1}
\bibliography{exact}
\end{document}